%% file: MasterDocument.tex
\documentclass[10pt,onecolumn, a4paper]{article}
\usepackage{times}
\usepackage{graphicx}
\usepackage{caption}
\usepackage{rotating}
\usepackage[authoryear]{natbib}
\usepackage[nofiglist, notablist]{endfloat}
\usepackage[]{efxmpl}
 \usepackage{setspace}
\usepackage{fancyhdr}
\newcommand{\Section}[1]{\vspace{-8pt}\section{\hskip -1em.~~#1}\vspace{-3pt}}

\makeatletter
\def\unnumfootnote{\xdef\@thefnmark{}\@footnotetext}

\begin{document}


\title{\Large\bf An Empirical Analysis of Dynamic Multiscale Hedging \\ using Wavelet Decomposition}

\author{Thomas Conlon and John Cotter}

\unnumfootnote{John Cotter, Director of Centre for Financial Markets, Department of Banking and Finance, University College Dublin, Blackrock, Co. Dublin, Ireland tel. +353 1 7168900, e-mail john.cotter@ucd.ie. Thomas Conlon, Postdoctoral Researcher, Financial Mathematics and Computation Research Cluster, Department of Banking and Finance, University College Dublin, Blackrock, Co. Dublin, Ireland tel. +353 1 7168909, e-mail conlon.thomas@ucd.ie. This publication has emanated from research conducted with the financial support of Science Foundation Ireland under Grant Number 08/SRC/FM1389.}

\maketitle

\thispagestyle{fancy}
\renewcommand{\headrulewidth}{0pt} 
\fancyhf{}
\rfoot{Paper to appear in \bf{\textsl{The Journal of Futures Markets}}} 

\section*{\centering Abstract}
{
This paper investigates the hedging effectiveness of a dynamic moving window OLS hedging model, formed using wavelet decomposed time-series. The wavelet transform is applied to calculate the appropriate dynamic minimum-variance hedge ratio for various hedging horizons for a number of assets.  The effectiveness of the dynamic multiscale hedging strategy is then tested, both in- and out-of-sample, using standard variance reduction and expanded to include a downside risk metric, the time horizon dependent Value-at-Risk.  Measured using variance reduction, the effectiveness converges to one at longer scales, while a measure of VaR reduction indicates a portion of residual risk remains at all scales.  Analysis of the hedge portfolio distributions indicate that this unhedged tail risk is related to excess portfolio kurtosis found at all scales.
}

\Section{Introduction}
The use of derivative securities, in particular futures contracts, allows both producers and consumers to reduce potential future price risk associated with a given spot position. Much of the large body of literature written on the issue of futures hedging have focussed either on the empirical estimation of the optimal hedge ratio (OHR) or the derivation of the OHR using different objective functions.  Many approaches to obtain optimal hedge ratios have been suggested, both static and dynamic.  Static hedging techniques include minimum variance, mean-variance, mean-Gini and generalized semi-variance.  Dynamic hedge ratios have also been proposed, applying techniques such as GARCH or moving-window estimation to capture changes in the relationship between assets.\footnote{Here, we follow the \citet{Chen2003} breakdown between static and dynamic hedging techniques.  These alternative methodologies are reviewed here and references therein.}  However, few empirical studies consider the effect of the hedging horizon on the optimal hedge ratio, even though various hedging participants may have very different hedging horizons.  Due to the sample reduction problem associated with matching the frequency of data with the hedging horizon, analysis of dynamic hedging at different time-horizons has been little studied.  In this paper, we overcome this difficulty by combining wavelet multiscale analysis with a moving window OLS, to calculate the time and scale dependent covariance structure and hence determine the dynamic time horizon dependent hedge ratio.  We then build further upon previous studies, by measuring the effectiveness at each time-horizon using a value-at-risk (VaR) measure, to assess the tail risk of the hedge portfolio at each scale.  Finally, to try to understand better the changes in effectiveness at different scales, we expand upon previous studies and explore the distributional characteristics of portfolio returns at different horizons by determining the scale dependent moments including skewness and kurtosis.  A number of implications for hedgers emerge from our findings.  First, hedgers with a longer time horizon benefit from lower levels of risk, higher effectiveness and lower transaction costs.  Second, static multiscale hedge ratios, found in previous studies, result in a smoothing of the data, which obscures the large dynamical changes that occur over time.  A dynamic multiscale method is shown to be more appropriate, capturing features not apparent using a static method.  Finally, while previous studies have demonstrated little hedge portfolio risk at longer scales, we find using a VaR effectiveness measure, that excess unhedged tail risk remains.  This highlights the weakness of the minimum variance hedge, even at long time-horizons.

The risk of financial assets is uniquely shaped by the time-horizon studied.  In the context of hedging, a limited number of studies, including \cite{Ederington1979, Hill1982, Malliaris1991,Benet1992, Geppert1995}, have demonstrated an increase in hedging effectiveness for longer horizons, by matching the frequency of the data with the hedging horizon.  However, out-of-sample, \cite{Malliaris1991, Benet1992} found a lack of stability in the hedging effectiveness for longer horizons.  More recently, \citet{Chen2004} demonstrated, using subsampled data, that both the hedge ratio and effectiveness tend to increase with the length of time horizon.\footnote{In this article, we define subsampled data as returns calculated from price data of a longer horizon, found by subsampling the original asset prices.  For example, one can create monthly returns from daily prices by subsampling the data every twenty days and calculating the return.  However, this has the obvious effect of reducing the sample size available, something we attempt to overcome in this study by using a wavelet approach.}  The effectiveness of scaled short-term horizon data applied to longer-term horizons was studied by \citet{Cotter2010a}, where scaled hedges were shown to provide good hedging effectiveness across a number of assets.   In all of these studies, the returns were calculated by sub-sampling over different horizons resulting in reduced quantities of data for longer-term horizons.  In order to overcome the sample reduction difficulties associated with reduced data quantity, we apply wavelet multiscaling techniques which allow us to compute the hedge ratio based on all data available at each scale.\footnote{It should be noted that these data points are based upon the same sample, which may result in a reduction of the precision at longer scales.}

Using wavelet multiscale analysis, we compute the hedge ratio and study hedging effectiveness at different time horizons.  Wavelets have previously been applied to a variety of economic and financial time series to decompose the data into orthogonal time-scale components of varying granularities.\footnote{Early applications included the study of foreign exchange data using waveform dictionaries, \citep{Ramsey1997}, the decomposition of economic relationships, \citep{Ramsey1998}, scaling properties of volatility, \citep{Gencay2001a} and the relationship between systematic risk and return at different scales, \citep{Genay2003}.  More recently, the relationship between stock returns and inflation, \citep{Kim2005}, the co-skewness and co-kurtosis between equities and the market at various time-scales, \citep{Galagedera2008a}, the scale dependence of hedge fund market risk and correlation, \citep{Conlon2008} and international diversification benefits at different time horizons, \citep{Rua2009} have been studied using the wavelet transform.}  Recently, wavelet multiscaling techniques have been applied to test the dependence of the futures hedge ratio on the underlying time-scale structure of the data.  By calculating the wavelet variance and covariance at different scales for S\&P 500 index and futures data, \cite{In2006a} showed that there is a unique hedge ratio associated with each scale, which converges to one for longer scales.  Further, using the level of variance reduction as a measure of hedging effectiveness, they demonstrated that the hedging effectiveness also converges to one.  Similar results were found between the Australian All Ordinaries Index and the Sydney Futures Exchange Share Price Index, \citep{In2006b}.

A comparison of wavelet multiscale hedge ratios to other approaches has also been addressed, \citep{Lien2007}.  Comparison to the error-correction hedge ratio revealed an outperformance for short time-horizons, while for long time-horizons, the optimal multiscale wavelet ratio was found to dominate, with similar results found both in- and out-of-sample.  The optimal hedge ratios for a portfolio of commodities was found, \cite{Fernandez2008}, using copulas to measure the asset returns dependency and wavelets to account for hedging horizon.  Improved hedging effectiveness was found for the portfolio of commodities compared to a single position, with additional benefits at longer scales.  While these previous studies detailed the effects of time-horizon on the hedge ratio, the relationship between the cash and futures is assumed to be static.  Assuming a static hedge ratio may restrict the introduction of newly available information which may impact the covariance structure and, hence, the hedge ratio.  In order to incorporate the characteristic time varying covariance, we expand upon these previous studies through the development of a dynamic multiscale hedge ratio.

It is well documented in the literature that the relationship between asset returns is time varying.  In this article, we follow the methods of \cite{Malliaris1991}, \cite{Harris2003} and \cite{Cotter2006a} and use a rolling window OLS, in order to capture changes in the covariance structure over time.  This method, combined with wavelet multiscaling, allows us to measure the hedging effectiveness both in- and out-of-sample for different time-horizons, providing a simultaneous time-scale measurement of multiscale hedging effectiveness.\footnote{Alternative approaches, not pursued here, to capture the time-varying covariance structure include GARCH models, \citep{Chen2003}.} Further, the wavelet multiscaling technique used is not subject to downsampling, (or reduction of the number of coefficients at longer scales), thus allowing us to align the hedge ratio and effectiveness features at different scales for dynamic comparison.

The performance of the hedging effectiveness at differing time-horizons is measured using two different methods, variance and value-at-risk reduction.  The variance method alone has been applied in previous wavelet multiscale hedge ratio studies and measures the reduction in hedge portfolio variance, compared to the unhedged.  However, variance assigns an equal weight to positive and negative returns, while a measure that differentiates between positive and negative returns may capture the hedger's preferences better.  In order to study the effect of hedging on the negative tail returns of the hedge portfolio, we also use value-at-risk (VaR) reduction, (see \cite{Cotter2006a,Harris2006,Cao2009}).\footnote{Alternatively, we could examine hedging for positive tail returns but, for conciseness, we illustrate the use of VaR as a performance evaluation method for a single side of the distribution}  When returns are normally distributed with mean zero, the VaR is simply a multiple of the standard deviation of the portfolio.  However, for non-normal returns, the VaR takes into account the higher moments of the distribution and so, improves upon the variance.  The second issue addressed in this article is the measurement of the effectiveness, at different scales, using both a standard variance metric and a VaR measure to explore scale dependent tail risks.\footnote{The semivariance was also tested as a measure of hedging effectiveness at each scale.  However, both in- and out-of-sample the effectiveness was found to be very similar to that of the variance measure.  These results are available on request.}

While a number of studies have shown a reduction in portfolio variance at longer time-horizons, the effect of time scale on the skewness and kurtosis, and hence the tail risk of a hedge portfolio, has not been examined.  The use of variance as a measure of risk is only correct when investors have a quadratic utility and returns are elliptically distributed, \citep{Harris2006}.  When these conditions do not hold, variance cannot characterize fully the risks associated with higher moments of the returns.\footnote{See \cite{Christie-David2001} where they demonstrate the importance of skewness and kurtosis in explaining the return-generating process of futures.}  Previously, \cite{Harris2006} found the skewness of the minimum-variance hedge portfolio to be little changed, while the portfolio kurtosis tended to increase compared to the unhedged asset, using original daily returns data.   Thus, the final issue addressed in this article is the effect of time-scale on the portfolio skewness and kurtosis and hence the risk of the hedged futures portfolio.

This paper is organized as follows.  In Section~\ref{Methods}, we describe the application of wavelets to decompose returns into component scales and then describe the optimal hedge ratio and hedging effectiveness measures.  Data and empirical results are described in Section~\ref{Results}, while some concluding remarks are given in Section~\ref{Conclusions}.

\section{Methodology}
\label{Methods}
\subsection{Wavelet Multiscale Analysis}
\label{DWT}
We provide a short synopsis of wavelet multiscale analysis relevant to this study, (for more comprehensive detail, see \cite{Burrus1997, Percival2000}).  The discrete wavelet transform provides an efficient means of studying multiresolution properties, as it can be used to decompose a signal into different time horizons or frequency components.  There are two basic wavelet functions, the father wavelet $\phi$ and mother wavelet $\psi$, which can be scaled and translated to form a basis for the Hilbert space $L^2 (\Re)$ of square integrable functions. The father and mother wavelets are formally defined by the functions:
\begin{eqnarray}
\phi_{j,k}\left(t\right) = 2^{-\frac{j}{2}} \phi\left(2^{j}t - k\right) \\
\psi_{j,k}\left(t\right) = 2^{-\frac{j}{2}} \psi\left(2^{j}t - k\right) 
\end{eqnarray}
where $j = 1, \ldots J$ is the scaling parameter in a $J$-level decomposition and $k$ is a translation parameter.  The long scale trend of the time series is captured by the father wavelet, which integrates to $1$, while the mother wavelet, which integrates to $0$, describes fluctuations from the trend.  The wavelet representation of a discrete signal $f(t)$ in $L^{2}(\Re)$ is given by:
\begin{eqnarray}
f(t) & = & \sum_{k}s_{J,k}\phi_{J,k}(t) + \sum_{k}d_{J,k}\phi_{J,k}(t)+ \ldots  +   \sum_{k}d_{1,k}\phi_{1,k}(t)  
\end{eqnarray}
where $k$ ranges from $1$ to the number of coefficients in the specified level and $J$ is the number of multiresolution levels, (scales). Smooth and detail component coefficients, $s_{J,k}$ and $d_{J,k}$, are found by integrating over time, $dt$,
\begin{eqnarray}
s_{J,k} = \int \phi_{J,k} f(t)dt &  	& \\
d_{j,k} = \int \psi_{j,k} f(t)dt & 	  &   	(j = 1, \ldots J)
\end{eqnarray}
Each coefficient sets $s_{J},d_J,d_{J-1}, \ldots d_1$ is called a \emph{crystal}, where coefficients from level $j = 1 \ldots J$ are associated with scale $[2^{j-1}, 2^{j}]$.

\subsubsection{MODWT}
\label{MODWT}
In order to overcome some of the difficulties associated with the DWT, in this paper we adopt the maximum overlap discrete wavelet transform (MODWT), a highly redundant linear filter that transforms a series into coefficients related to variations over a set of scales, \citep{Percival2000,Gencay2001b}.  The MODWT has several advantages over the DWT, allowing alignment of wavelet scaling and detail coefficients with the original time-series.  The MODWT can also handle any sample size N, whereas the DWT restricts the sample size to a multiple of $2^{j}$.  Here, we apply the MODWT as it allows us to explore any sample size, align the coefficients with the original data and calculate the wavelet variance and covariance effectively at different scales.

Like the DWT, the MOWDT produces a set of time-dependent wavelet and scaling coefficients with basis vectors associated with a location $t$ and scale $\tau_{j} = [2^{j-1}, 2^{j}]$ for each decomposition level $j = 1, \ldots, J_{0}$. However, the MODWT is nonorthogonal and has a high level of redundancy, retaining downsampled values at each level of the decomposition that would be discarded by the DWT.\footnote{Downsampling or decimation of the wavelet coefficients retains half of the number of coefficients that were retained at the previous scale and is applied in the Discrete Wavelet Transform. By retaining all coefficients at each scale, the MODWT has 'redundant' coefficients or coefficients not necessary to recreate the original signal.  This, however, results in significant benefits for multiscale analysis, described above.} Decomposing a signal using the MODWT to $J$ levels theoretically involves the application of $J$ pairs of filters.  The filtering operation at the $j^{th}$ level consists of applying a rescaled father wavelet to yield a set of \emph{detail coefficients}
\begin{equation}
\tilde{D}_{j,t} = \sum^{L_{j}-1}_{l = 0} \tilde{\psi}_{j,l} f_{t-l}
\end{equation}
and a rescaled mother wavelet to yield a set of \emph{scaling coefficients}
\begin{equation}
\tilde{S}_{j,t} = \sum^{L_{j}-1}_{l = 0} \tilde{\phi}_{j,l} f_{t-l}
\end{equation}
for all times $t = \ldots,-1,0,1,\ldots$, where $f$ is the function to be decomposed, \citep{Percival2000}.  The rescaled mother, $\tilde{\psi}_{j,l}=\frac{{\psi}_{j,l}}{2^{j}}$, and father, $\tilde{\phi}_{j,t} = \frac{{\phi}_{j,l}}{2^{j}}$, wavelets for the $j^{th}$ level are a set of scale-dependent localized differencing and averaging operators and can be regarded as rescaled versions of the originals.  The $j^{th}$ level equivalent filter coefficients have a width $L_{j} = (2^j - 1)(L - 1) + 1$, where $L$ is the width of the $j=1$ base filter.  In practice, the filters for $j>1$ are not explicitly constructed because the detail and scaling coefficients can be calculated, using an algorithm that involves the $j=1$ filters operating recurrently on the $j^{th}$ level scaling coefficients, to generate the $j+1$ level scaling and detail coefficients, \citep{Percival2000}.

\subsubsection{Wavelet Moments and Covariance}
The wavelet variance $Variance_{f} (\tau_j)$ at scale $j$ is defined as the expected value of $\tilde{D}^{2}_{j,t}$ if we consider only the non-boundary coefficients.\footnote{The MODWT treats the time-series as if it were periodic using ``circular boundary conditions''.  There are $L_j$ wavelet and scaling coefficients that are influenced by the extension, which are referred to as the boundary coefficients.}  An \emph{unbiased} estimator of the wavelet variance for function $f(t)$ at scale $j$ is formed by removing all coefficients that are affected by boundary conditions and given by:
\begin{equation}
Variance_{f} (\tau_j) = \frac{1}{M_j} \sum^{N-1}_{t = L_{j}-1} \tilde{D}^{2}_{j,t}
\label{W_var}
\end{equation}
where $M_j = N - L_j +1$ is the number of non-boundary coefficients at the $j^{th}$ level associated with the time horizon $\tau$, \cite{Percival2000}.  The wavelet variance decomposes the variance of a process on a scale-by-scale basis (at increasingly higher resolutions of the signal) and allows us to explore how a signal behaves at different time horizons. 

Similarly, the wavelet skewness and kurtosis can be defined on a scale-by-scale basis.  Assuming that the wavelet coefficients $D_{j,t}$ at each scale have zero mean, the unbiased skewness at each scale is given by
\begin{eqnarray}
Skewness_{f} (\tau_j)  & = & \frac{ \frac{1}{M_j} \sum^{N-1}_{t = L_{j}-1} \tilde{D}^{3}_{j,t} }{\sigma_f(\tilde{D}_{j,t})^{3}} 
\label{W_skew}
\end{eqnarray}
while the unbiased kurtosis at each scale is
\begin{eqnarray}
Kurtosis_{f} (\tau_j) & = & \frac{ \frac{1}{M_j} \sum^{N-1}_{t = L_{j}-1} \tilde{D}^{4}_{j,t} }{\sigma_f(\tilde{D}_{j,t})^{4}} 
\label{W_kurt}
\end{eqnarray}
with $\sigma^2_f (\tau_j)= {Variance_{f}(\tau_j)}$ the standard deviation of the wavelet coefficients at scale $j$.  Similarly, formulas for the co-skewness and co-kurtosis at different scales have been derived, \citep{Galagedera2008a}.  

In our analysis, we use (\ref{W_skew}) and (\ref{W_kurt}), to examine the higher moments of the hedge portfolio, in order to gain insight into the distributional effects of scaling. As described in \cite{Harris2006}, the skewness and kurtosis of the minimum variance hedge portfolio can be larger than that of the individual assets, creating a need for a measure of risk that captures these higher moments.   Here, we also address the question of how time horizon effects the higher moments of the hedge portfolio and hence, the tail risk of the hedge portfolio.

The wavelet covariance between functions $f(t)$ and $g(t)$ is defined, similar to (\ref{W_var}), to be the covariance of the wavelet coefficients at a given scale.  The \emph{unbiased} estimator of wavelet covariance at the $j^{th}$ scale is given by
\begin{equation}
Covariance_{fg} (\tau_j) = \frac{1}{M_j} \sum^{N-1}_{t = L_{j}-1} \tilde{D}_{j,t}^{f(t)} \tilde{D}_{j,t}^{g(t)}
\label{W_covar}
\end{equation}
where all wavelet coefficients affected by the boundary are removed and $M_j = N - L_j +1$, (see \cite{Percival2000} for a complete treatment of wavelet moments).

\subsection{Minimum Variance Hedge}
In this paper, we use the wavelet transform so as to calculate the minimum variance hedge ratio at different time horizons.  For an individual holding a spot position in some asset, hedging involves taking an opposite position in the futures market.   Assuming a long position in the spot market, the return on a hedge portfolio is given by
\begin{equation}
r_t = s_t - hf_t
\end{equation}
where $f_t$ and $s_t$ are the log returns of the futures and spot markets at time $t$ and $h$ is the hedge ratio.  The risk of a portfolio, commonly given as the variance in returns is
\begin{eqnarray}
Var(r_t) & = & Var(s_t - hf_t) \nonumber \\
& = & Var(s_t) + h^{2}Var(f_t) - 2h Cov(s_t,f_t)
\label{Var_Returns}
\end{eqnarray}
The static minimum variance hedge ratio is the value of $h$ that minimizes (\ref{Var_Returns}), and given by
\begin{equation}
h =  \frac{Coviance(s,f)}{Variance(f)}
\label{OHR}
\end{equation}
where $Covariance(sf)$ is the covariance between the spot and futures returns and $Variance(f)$ the variance of the futures returns.  However, the time variation of the variance-covariance matrix is a well known feature of many financial asset returns leading to an optimal time-dependent hedge ratio, $h_t$, \citep{Kroner1993},
\begin{equation}
h_t =  \frac{Coviance(s_t,f_t)}{Variance(f_t)}
\label{OHR_t}
\end{equation}
In order to account for this time-variation in the variance-covariance matrix, a rolling window OLS approach is used, with all observations given an equal weighting. This approach combines well with the wavelet transform, allowing a dynamic scale dependent analysis of the hedge ratio.  To calculate the hedge ratio at each scale, we simply replace the variance and covariance in (\ref{OHR_t}) by that found using the wavelet coefficients at each scale for each moving window, (\ref{W_var}) and (\ref{W_covar}).


\subsection{Hedging Effectiveness}
We examine both in-sample and out-of-sample hedging performance using two different performance metrics, variance reduction, which incorporates both upside and downside risk, and value-at-risk reduction which captures risk for one side of the distribution in our case.

Variance reduction measures the percentage reduction in the variance of a hedge portfolio compared to the unhedged spot position and is given by
\begin{equation}
HE_{variance} = 1 - \frac{Variance(r_t)}{Variance(s_t)}
\label{HE_variance}
\end{equation}
where $Variance(r_t)$ and $Variance(s_t)$ are variance of the returns for the hedge portfolio and spot respectively.

To study the effect of hedging on negative tail returns and measure risks posed by higher moments of portfolio returns, we use Value-at-Risk (VaR), which estimates the maximum portfolio expected loss for a given confidence level over a given time period, (\cite{Jorion2006,Harris2006}).  The VaR at confidence level $\alpha$ is
\begin{equation}
VaR_{\alpha} = q_a
\end{equation}
where $q_a$ is the relevant quantile of the loss distribution.  The effectiveness from the point of view of Value-at-Risk reduction is then measured by
\begin{eqnarray}
HE_{VaR} = 1 - \frac{VaR_{\alpha}(r_t)}{VaR_{\alpha}(s_t)}
\end{eqnarray}
with $VaR_{\alpha}(r_t)$ and $VaR_{\alpha}(s_t)$ the value-at-risk of the hedge portfolio and spot, at confidence level $\alpha$.

\section{Empirical Analysis}
\label{Results}
\subsection{Data}
For the empirical analysis, we choose examples of three asset classes.  We dynamically hedge long spot exposures in West Texas Intermediate (WTI) Crude Oil, the S\&P $500$ Equity Index and the $GBP/USD$ currency exchange rate.  These assets were chosen to represent a diverse set of highly liquid cash and futures markets, where a long returns history for both the spot and futures markets was available.\footnote{Additional assets, (eg. Gold), were also studied and the results found to be consistent.  These are not presented for conciseness.} Each long spot position is hedged by taking a short position in the corresponding futures contract.   The empirical results are found using daily returns data for the period $02$ Jan $1986$ to $31$ December $2009$:

\begin{enumerate}
\item
For WTI Crude Oil, the corresponding futures contract is the New York Mercantile Exchange (NYMEX) contract, giving a total of $6259$ trading days.
\item
The S\&P $500$ data consists of $6238$ daily traded returns, with the futures contract traded on the Chicago Mercantile Exchange.
\item
The British Pound to US Dollar exchange rate, with a total of $6261$ daily returns and the futures contract is traded on the Chicago Mercantile Exchange.

\end{enumerate}
The timeframe examined was chosen as it covered a large number of different adverse events for each asset, allowing a detailed study of the effects of scaling on hedging effectiveness, during both normal and turbulent markets.  Data was obtained from Datastream, using closing prices for the spot index and the corresponding daily settlement price for the futures contract.  Each futures contract studied is nearest-to-maturity and rolled over to the next contract on the first day of the contract month.  

As outlined in Section~\ref{MODWT} we decompose both the cash and futures returns by employing the MODWT.  For the present study, we selected the least asymmetric (LA) wavelet, (known as the Symmlet, \citep{Burrus1997}), chosen as it exhibits near symmetry about the filter midpoint and has the property of aligning the wavelet coefficients accurately with the unfiltered time series.\footnote{The Daubauchies $D4$ and the Coiflet $C10$ wavelets were also studied but resulted in little qualitative difference to the analysis.}  LA filters are defined in even widths and the optimal filter width is dependent on the characteristics of the signal and the length of the data series.  The filter width chosen for this study was the LA8, (where $8$ refers to the width of the scaling function).  The length of the rolling window used in the analysis is $1000$ days\footnote{Different window sizes were also studied, with longer windows found to smooth changes in the hedge ratio, while shorter windows resulted in more volatile ratios. However, the main results found in this paper were qualitatively the same regardless of the window studied.} and we chose scale $6$, corresponding to $32-64$ day dynamics, as the largest decomposition level, to strike a balance between the maximum scale and the number of boundary coefficients.  As described in Section~\ref{DWT}, the scales studied can be interpreted as follows:  Scale $1\rightarrow 1-2$ day, Scale $2\rightarrow 2-4$ day, Scale $3\rightarrow 4-8$ day, Scale $4\rightarrow 8-16$ day, Scale $5\rightarrow 16-32$ day and Scale $6\rightarrow 32-64$ day dynamics.

Summary statistics, including the mean, standard deviation, skewness, and kurtosis for each of the assets at each scale can be found in Table~\ref{Table1}.  Starting with the original returns data, we find the common stylized features of financial returns, namely, excess kurtosis and a lack of normality for both spot and futures.  Turning to the scale statistics, as described in Section~\ref{DWT}, the mother wavelet integrates to zero, and so the mean value for the wavelet decomposed data at each scale is zero.  As found in previous studies, the standard deviation decreases at lower scales, with total variance conserved.  The skewness is found to be predominantly negative across assets and scales, while excess kurtosis is found at all scales.  However, the level of kurtosis is found to decrease significantly at long scales.  The hypothesis of normality for the returns coefficients associated with each scale is found, however, to be rejected by the Jacque-Bera statistic for all assets.

\input{Table1.tex}

\subsection{Dynamic Hedging With Subsampled Data}
In this paper, we use wavelet multiscaling techniques to study the dynamic scale dependent hedge ratio, in order to overcome the sample reduction problems associated with sub-sampling data. To demonstrate the sub-sampling problems in the case of Crude Oil, we first study the changes in the optimal dynamic hedge ratio (and associated hedging effectiveness) for a number of sub-sampled time-horizons.\footnote{Results were found to be consistent for the other assets studied but are not reported for brevity.}  This is achieved by calculating asset returns from price data sampled every $3$, $6$ and $12$ days, reducing the quantities of data available for analysis. This ignores any information contained in the unused data, a problem that is overcome using wavelet multiscaling.  To allow for comparison with the rolling window wavelet techniques later, we estimate the returns in a rolling window of $200$ days, calculate the hedge ratio and then move forward one time-period, (dropping the first observation).\footnote{Longer windows reduced the available data further and would impair a detailed rolling window analysis.}

The results, averaged over each moving window for the time-horizons described are shown in Table~\ref{Subsampled_table}. As expected, an increase in the hedge ratio is found at longer time-horizons, with a corresponding increase in the hedging effectiveness.  The effects on skewness and kurtosis are more difficult to determine, with differing trends across scales, although the kurtosis using $3$, $6$ and $12$ day returns is lower than for the original daily returns. However, as previously shown by \cite{Harris2006} for cross-hedged currency portfolios, the kurtosis of the hedge portfolio was found to be greater than that of the spot for all returns horizons.

\begin{figure}[htbp!]
\begin{center}
\includegraphics[height=120mm,width=155mm]{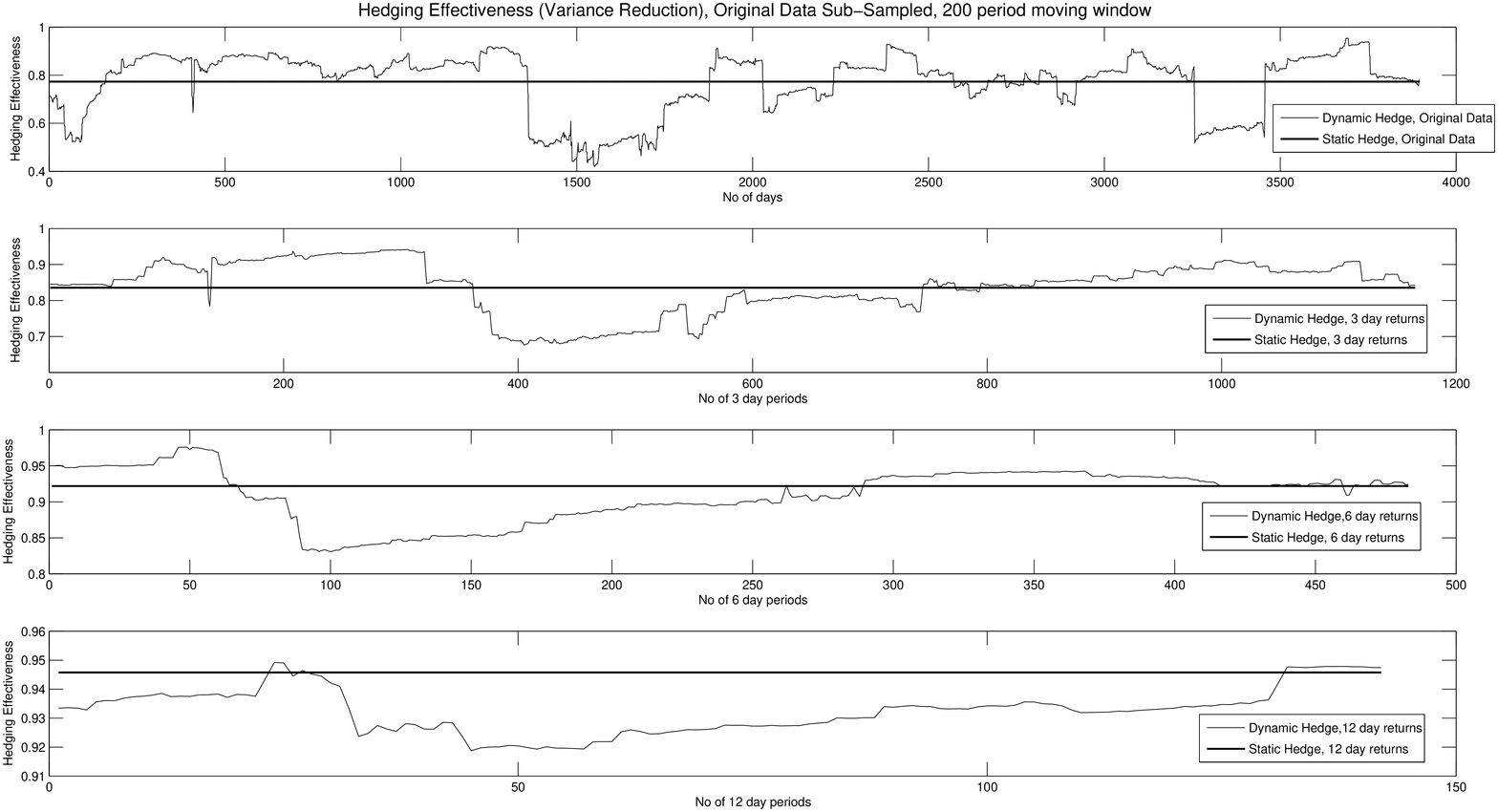}
\caption{\bf Dynamic hedging effectiveness \\ \rm Dynamic variance reduction hedging effectiveness is presented, in-sample, for a long Crude Oil position hedged with futures, for sub-sampled returns data with various time horizons, calculated using a rolling window of $200$ days.  The effectiveness improves at longer time horizons, however the number of time periods available for analysis decreases considerably impairing comparison between different horizons.}
\label{HE_ss}
\end{center}
\end{figure}

The difficulties of using sub-sampled data can be seen more clearly in Figure~\ref{HE_ss}, where in-sample hedging effectiveness, measured using variance reduction, for each time-window at each hedging horizon is shown.  The dynamic nature of hedging effectiveness is clearly visible, with quite dramatic variations using the original data.  The increased effectiveness at longer horizons is visible; however the reduction in sample data makes it difficult to match and compare features for different scales at any given point.  Further, the reduced quantities of available data prevents analysis at scales much longer than $12$ days as the statistical quality of the data deteriorates quickly, while also preventing a sensible out-of-sample analysis.  To attempt to overcome these difficulties, we apply wavelet decomposition.

\input{Table2.tex}

\subsection{Dynamic Scale Dependent Hedging}
In order to investigate the effects of both time and scale on the hedge ratio and hence the hedging effectiveness, we calculate the optimal dynamic hedge ratio using a moving-window technique.  This was implemented as follows: The wavelet coefficients for both spot and futures returns were calculated, up to the sixth scale, using a moving window of $1000$ days, allowing the variance, (\ref{W_var}), covariance, (\ref{W_covar}), and minimum variance hedge ratio, (\ref{OHR_t}), to be found for each scale in each window.  The in-sample hedging effectiveness at each scale was determined using the wavelet coefficients from the first $1000$ days.  The out-of-sample effectiveness at each scale was measured by applying the in-sample hedge ratio to the wavelet coefficients calculated over the next $1000$ days, (following the analysis of \cite{Benet1992,Chen2004} using subsampled data and \cite{Lien2007,Fernandez2008} using wavelet filtered data).  Thereafter, the observation at $T+1$ is incorporated into the data and the first observation excluded, with the above process repeated.

\begin{sidewaysfigure}[htbp!]
\begin{center}
\includegraphics[height=150mm,width=210mm]{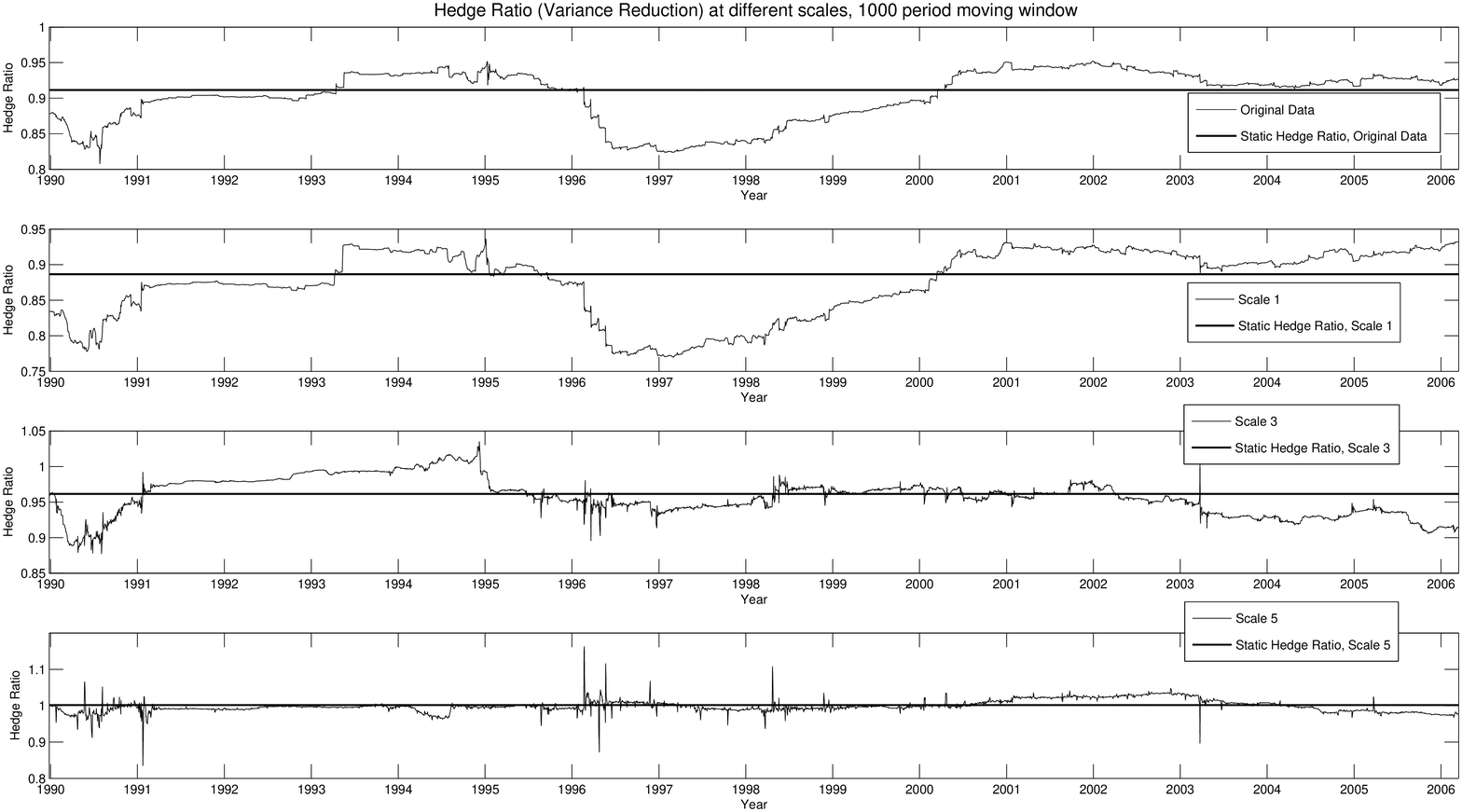}
\caption{\bf Crude Oil: Dynamic multiscale hedge ratio.  \\ \rm Notes: The dynamic minimum variance hedge ratio for Crude Oil was found using a rolling window of $1000$ days. The static hedge ratio, calculated using all available data is also shown.  At short scales, the dynamic hedge ratio is found to vary considerably over time, while at longer scales, although there are a number of spikes, the ratio converges to one.}
\label{HR}
\end{center}
\end{sidewaysfigure}

To illustrate, the dynamic hedge ratio for Crude Oil is shown, Figure~\ref{HR}, with the dynamic minimum variance hedge ratio for the original data shown in the upper plot, while those found at wavelet scales $1$, $3$ and $5$ are shown in the lower plots.\footnote{For brevity, the plots of the moving window analysis are not shown for the other assets.  However, summary statistics are shown in Table \ref{Table4} and Table \ref{Table5}, while the plots are available from the authors upon request.}   We find that the hedge ratio tends to one as we move to longer time scales.   However, we find that the ratio is far from static, in particular at short time-scales.   The dynamics of the ratio at scale $1$, ($1-2$ days), has a trend similar to that of the original data, although the value of the ratio is reduced somewhat.  By scale $5$, the ratio has converged to one; however there is evidence of some spikes in the data, which may be a result of considerable basis risk at that point.\footnote{For example, on $28$th January $1991$, coinciding with the end of the First Gulf War, spot Crude Oil fell in price by $6.0$\%, while the futures fell by only $0.8$\%.} The less dynamic nature of the hedge ratio at longer scales would have a major impact on the level of transaction costs involved for hedgers with a long term horizon.  For a hedger with a horizon of $16-32$ days and above (scale $5$ and above), the convergence of the hedge ratio to one means a consistent hedge, reducing the considerable transaction costs associated with the changes in the hedge ratio at shorter horizons.

The hedging effectiveness, measured using variance reduction, (\ref{HE_variance}), for each moving window is shown in Figure~\ref{HE_VaR}, with both the in- and out-of-sample effectiveness overlaid for comparative purposes.  There are a number of interesting points to note here.  First, the in-sample and out-of-sample effectiveness tend to be very close at all scales, indicating favourable performance of the dynamic multiscale hedging.  This is in contrast to the results found by \cite{Lien2007}, where the out-of-sample hedging effectiveness for Crude Oil was shown to improve relative to the in-sample at longer scales, only for a single window.  Second, moving to longer time scales, the variance reduction effectiveness (both in- and out-of-sample) increases significantly and by scale 6, ($32-64$ days), the effectiveness has, on average converged to one, (see Table~\ref{Table3}).  However, it must be noted that there occurs a number of singularities where the hedging effectiveness drops considerably at long scales.  Singularities such as these were not witnessed in previous multiscale hedging studies, as they examined the multiscaled properties of the entire dataset, effectively smoothly out these events but ignoring the dynamic nature of variance and covariance.  These large drops in effectiveness occur at days with large basis risk, demonstrating that even with a long hedging horizon, a hedger may be subject to basis risk.

\begin{sidewaysfigure}[htbp!]
\begin{center}
\includegraphics[height=150mm,width=210mm]{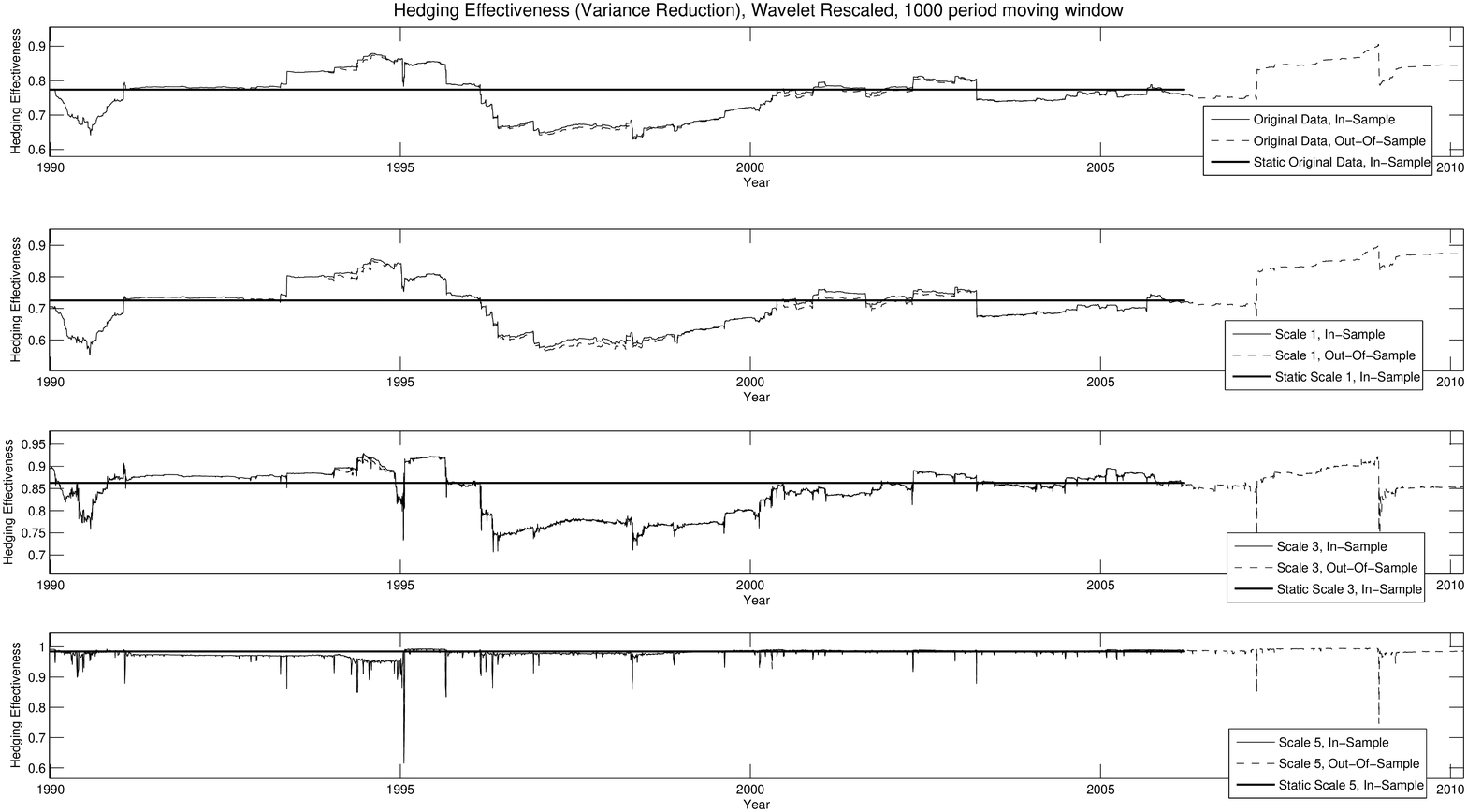}
\caption{\bf Crude Oil: Dynamic multiscale hedging effectiveness. \\ \rm Notes:  Hedging effectiveness, measured in terms of variance reduction, was found using a rolling window of $1000$ days with in- and out-of-sample results overlaid.  In- and out-of-sample results are found to track closely at all scales, while the effectiveness is found to increase at longer scales, with less variation.}
\label{HE}
\end{center}
\end{sidewaysfigure}

We now consider the effect of the minimum-variance hedge ratio on the $95\%$ value-at-risk of the hedge portfolio in Figure~\ref{HE_VaR}.  Similar to the analysis for variance reduction, we find that in- and out-of-sample effectiveness track each other closely.  Also, the hedging effectiveness was found to increase at longer scales.  However, at scale $5$ ($16-32$ days), the average VaR hedging effectiveness was $88\%$, compared to $98\%$ for the variance reduction measure, resulting from residual unhedged tail risk.  In fact, across all scales, VaR effectiveness was found to be weaker compared to the variance reduction measure.  This reduced effectiveness is due to the use of the minimum variance hedge, which considers only the second moment of the returns distribution.  As shown in Table \ref{Table1}, asset returns at all scales are non-normal, resulting from fat tails of the returns distributions.  To determine if higher moments are the cause of the differences between effectiveness measures, we consider the skewness and kurtosis of both the hedged and unhedged portfolios at each scale.  

\begin{sidewaysfigure}[htbp!]
\begin{center}
\includegraphics[height=150mm,width=210mm]{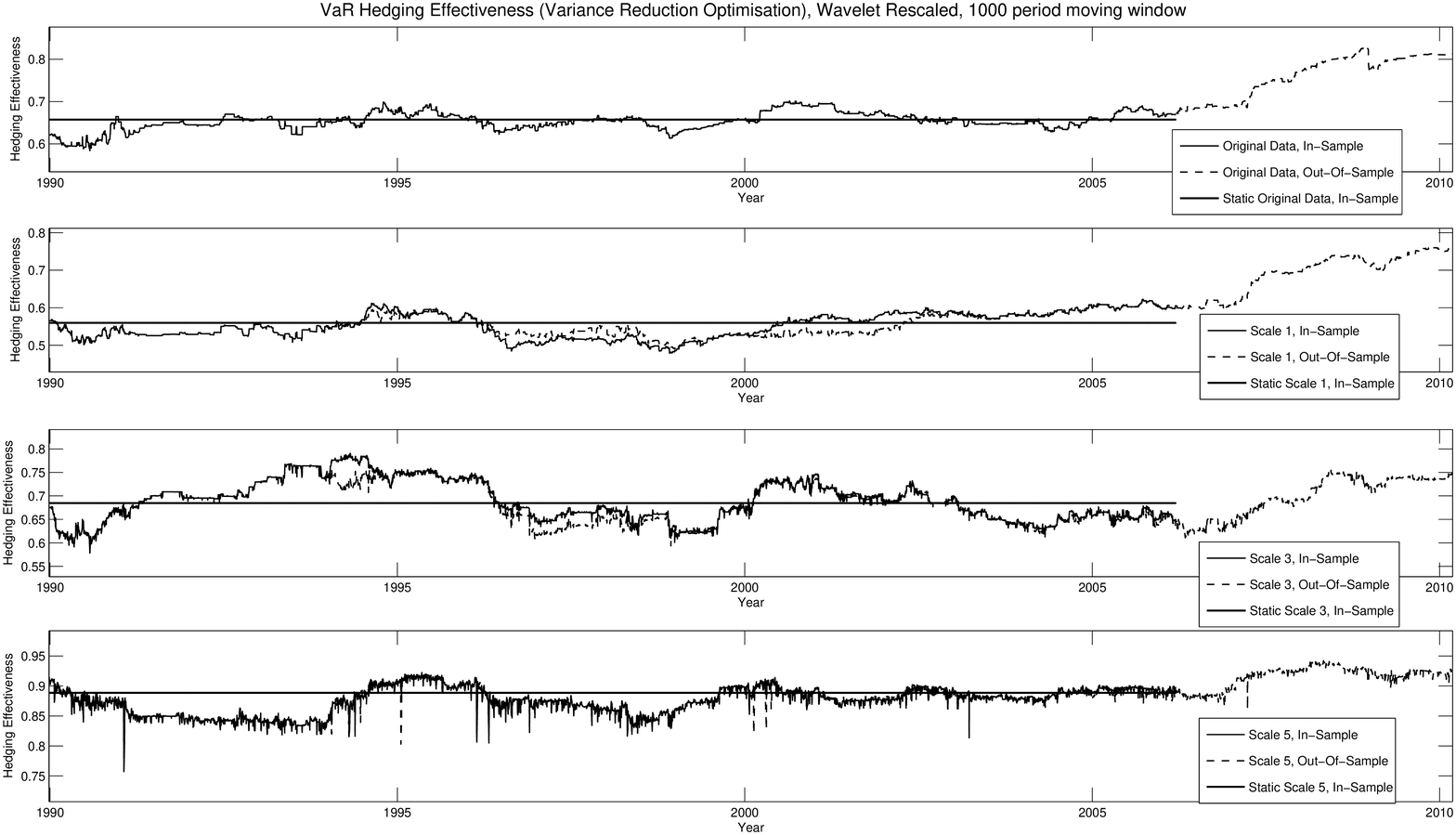}
\caption{\bf Crude Oil: Dynamic multiscale value-at-risk hedging effectiveness \\ \rm Notes: Hedging effectiveness, measured using VaR reduction, was calculated using a rolling window of $1000$ days with in- and out-of-sample results overlaid.  In- and out-of-sample results are found to track closely at all scales, while the effectiveness is found to increase at longer scales.  However, compared to the variance reduction effectiveness measure, (Figure \ref{HE}), some residual levels of risk remain even at the longest scale studied.}
\label{HE_VaR}
\end{center}
\end{sidewaysfigure}
Table~\ref{Table3}, displays summary statistics across windows and shows the average hedge ratio, hedging effectiveness ($95\%$ VaR and variance reduction), and the standard deviation, skewness and kurtosis of both the unhedged asset and the hedge portfolio at each scale for WTI Crude Oil.\footnote{The averages were found across each overlaid moving window using only data where both in- and out-of-sample results were available concurrently. This allows a direct comparison between the in- and out-of-sample results.}   These results are shown both in- and out-of-sample and, in both cases, we find that the standard deviation is reduced for both the unhedged and hedged portfolios at longer scales.  The skewness although predominantly negative, is more ambiguous across scales, with no distinct trend found.  However, considering the kurtosis, we find it is greater for the hedge portfolio across all scales, similar to that found by \cite{Harris2006} for daily returns data.  Also, for both hedged and unhedged portfolios, the level of kurtosis drops consistently as we move to longer scales.   However, even at the longest scale studied, we find excess kurtosis for the hedge portfolio, (while the unhedged portfolio has zero excess kurtosis).\footnote{This is in contrast to the findings of \cite{Galagedera2008a}, where the MODWT was used to demonstrate the excess kurtosis of a portfolio of equities to be, on average, positive at short scales while consistently negative at longer scales.}  This is in keeping with the findings of \cite{Harris2006} using daily unfiltered data and this may indicate that the excess kurtosis is behind the reduced effectiveness of the hedge portfolio from a value at risk perspective, (compared with the variance reduction measure).  This indicates that a technique that explicitly accounts for higher moments, such as VaR minimisation, see \cite{Harris2006,Cao2009}, may be more appropriate in reducing the risk, even at longer time horizons.

\input{Table3.tex}

Results, averaged across each moving window, for a minimum variance hedge portfolio consisting of a long position in the S\&P $500$ hedged with a short index futures position, are shown in Table~\ref{Table4} for different time scales.  Similar to that seen for Crude Oil, both the hedge ratio and the hedging effectiveness tend to increase at longer scales, (both in- and out-of-sample), although the VaR effectiveness tends to be lower than that measured by variance reduction.  Comparing the in-sample and out-of-sample results, we find that the dynamic minimum variance hedge ratio has slightly better in-sample effectiveness, although the small differences indicate the robustness of the method.  Examining the skewness of the returns for both the unhedged and hedged portfolio, again no distinct trend emerges across scales, although the hedged portfolio is found to always have negative skewness, in contrast to the unhedged asset.  Finally, analysing the level of kurtosis, we find limited situations where the kurtosis of the hedge portfolio is less than that of the unhedged portfolio, (at scales $1$ and $6$), something not witnessed for the other assets studied.  For the unhedged asset we find the kurtosis at scale one to be greater than that using the unfiltered original data, indicating that this time-scale picks up some large tail risks, (or large amplitude noise), not common to other scales.  As described by \cite{Benet1992}, this may indicate presence of a large amount of price uncertainty in this market for short time scales, while at longer timescales more information reduces the amount of uncertainty and hence the level of basis risk.  However, similar to the other assets we find kurtosis decreases at longer scales for both the unhedged and hedged assets. As the VaR hedging effectiveness is only $0.87$ at the longest scale, this suggests that higher order moments may also have an influence on the tail risk of the portfolio, (again, a VaR minimisation technique might help in reducing or eliminating these risks).

\input{Table4.tex}

The final dataset examined is a British Pound/US Dollar spot position hedged using futures, with results averaged over each $1000$ day moving window, shown in Table~\ref{Table5}.  As found previously for other assets, the hedging effectiveness was similar in- and out-of-sample across all scales, indicating the robustness of the dynamic method for futures hedging.  Similar to previous assets, we also find that the hedge ratio and effectiveness at the first scale, ($1-2$ day horizon), is substantially reduced compared to the other scales and to the original data.  This suggests that there is more market uncertainty at short scales or in a statistical sense that there is greater noise, increasing the difficulty of measuring the hedge ratio accurately, \cite{Benet1992}.  However, by the second scale ($2-4$ day horizon) the effectiveness is greater than that found using the original data, and then increases to a maximum of $0.98$ (variance reduction) or $0.88$ (VaR reduction) at long scales.  Examining the skewness, we find that the unhedged portfolio skewness turns negative at the fourth scale, while for the hedge portfolio it remains positive for the majority of scales.   Finally, the kurtosis of the hedge portfolio is found to be greater than the unhedged across all scales, with both decreasing at larger scales.  However, even at the largest scale, excess kurtosis persists suggesting that variance minimisation may not eliminate all portfolio risk, \citep{Harris2006}.  This is substantiated by the residual portfolio tail risk found using VaR effectiveness, even at long scales.

\input{Table5.tex}

Common to all the data sets studied is a reduction in the standard deviation of the dynamic hedge ratio at longer scales.   Compared to an agent with a short time-horizon, one with a longer horizon can substantially reduce the transaction costs involved, further enhancing the benefits of long-horizon hedging.  This results from the convergence of dynamic hedging to an almost static hedging strategy at long horizons producing, as demonstrated, improved levels of risk management.  VaR hedging effectiveness was found, for all assets across scales, to be less than that measured using variance reduction.  Similarly, the kurtosis was found to be larger for the hedge portfolio at all scales, while both the hedge and unhedged kurtosis decrease at longer scales.  By incorporating dynamic changes in covariance, we have demonstrated the benefits of a time varying multiscale hedge ratio, compared to the static multiscale approach considered in previous studies.

\section{Conclusions}
\label{Conclusions}
In this study, we apply the wavelet transform to investigate multiscale properties of both the hedge ratio and effectiveness of a futures hedge in a dynamic framework. We extend previous work by combining a moving-window OLS with wavelet decomposition in order to examine the time-scale behaviour dynamically.  By calculating the minimum variance optimal hedge ratio at different wavelet scales in each window, we examine both the in- and out-of-sample effectiveness.  Studying the results over the different moving-windows, we demonstrate the effectiveness of the dynamic method through the close tracking of the in- and out-of-sample hedging effectiveness at all scales.  The scale dependence of the hedge ratio and the convergence to one for longer time-horizons are also shown, with a reduction in the standard deviation of the hedge ratio at longer scales, (leading to reduced transaction costs for a hedger with long horizon).

Hedging effectiveness is measured first by calculating the fraction of the unhedged portfolio variance removed by hedging.  However, the variance measures only the second moment of the returns distribution and may not capture rare negative tail returns.  To test the hedging performance in the negative tail of the returns, we also measure the fraction of the $95$\% Value-at-Risk of the unhedged portfolio removed by hedging.  For both measures of hedging performance, the effectiveness is found to increase for longer time-horizons both in- and out-of-sample, with the variance reduction measure converging to one for all assets.  However, measured using Value-at-Risk, the effectiveness, although increasing does not converge to one at the longest horizons studied.  Thus, the application of variance minimisation to find the optimal hedge ratio, minimizes the portfolio variance but ignores higher moments, resulting in excess residual tail risk for the hedge portfolio even at long time horizons.

To investigate further the effects of minimum variance hedging at different scales, we examine returns distribution at all scales.   The skewness of hedge portfolio returns has little consistency across assets or scales.  However, the kurtosis for both the hedged and unhedged portfolios decreases as the hedging horizon increases, reducing the levels of tail risk, (as evidenced by the improvement in the VaR effectiveness measure).  The portfolio kurtosis is, on average, greater than the unhedged asset.  For both Crude Oil and British Pound/US dollar hedges, the hedge portfolio has excess kurtosis at all scales perhaps contributing to the extra tail risk found using VaR effectiveness.   

The implications of our findings are as follows:  Hedgers with a longer time horizon benefit from lower levels of risk, higher effectiveness and lower transaction costs.  The static multiscale hedge ratios, found in previous studies, result in a smoothing of the data, which obscures the large dynamical changes that occur over time.  A dynamic multiscale method is shown to be more appropriate, capturing features not apparent using a static method.  Additionally, while previous studies have demonstrated little hedge portfolio risk at longer scales, we find using a VaR measure, that excess unhedged tail risk remains.  This highlights the weakness of the minimum variance hedge, even at long time-horizons.

\bibliographystyle{authordate1new}
\bibliography{bibliography}

\end{document}

%% file: Table1.tex
\begin{sidewaystable}[htbp!]
\begin{center}
	\begin{tabular}{c || c c c c c | c c c c c c}
	
	\multicolumn{11}{c}{\bf{Crude Oil}} \\
		& \multicolumn{5}{c}{Cash} &				\multicolumn{5}{c}{Futures}				\\
			& Mean	& Standard	& Skewness & Kurtosis	& Jacque-Bera & Mean	 & Standard	& Skewness &	Kurtosis  & Jacque-Bera\\
			& 	& Deviation	&  &  & 	 &  & Deviation	& & &	 \\
			& \% 	& \%& & 	 &   & \%	& \% &	 \\
			\hline
		Original Data	& 0.0191	& 2.65 & -0.79	& 17.42	& 53046 & 0.0191	& 2.58	& -0.82	& 17.33 & 54283\\
		Scale 1	&	0	&	1.88	&	-0.31	&	11.70	& 19860 &	0	&	1.83	&	-0.19	&	11.54 & 19080 \\
		Scale 2	&	0	&	1.35	&	-0.30	&	15.31 & 39604	&	0	&	1.33	&	-0.26	&	16.51 & 47674 \\
		Scale 3	&	0	&	0.97	&	-0.12	&	9.45	& 10880 &	0	&	0.93	&	-0.07	&	7.99 & 6509 \\
		Scale 4	&	0	&	0.63	&	0.02	&	6.15	& 2581 &	0	&	0.60	&	-0.05	&	5.17 & 1235 \\
		Scale 5	&	0	&	0.38	&	-0.05	&	3.67 & 118	&	0	&	0.37	&	-0.06	&	3.57 & 88 \\
		Scale 6	&	0	&	0.27	&	-0.07	&	3.62 & 104	&	0	&	0.27	&	-0.07	&	3.63 & 108 \\
		\multicolumn{11}{c}{\bf{S\&P 500}} \\
		& \multicolumn{5}{c}{Cash}  & 			\multicolumn{5}{c}{Futures}			\\
			& Mean	& Standard	& Skewness & Kurtosis	& Jacque-Bera & Mean	 & Standard	& Skewness &	Kurtosis  & Jacque-Bera\\
			& 	& Deviation	&  & 	& &  & Deviation	& & &	 \\
			& \% 	& \%& & 	 &   & \%	& \% &	 \\
		\hline		
		Original Data	&	0.0267	&	1.18	&	-1.39	&	33.61	& 245512 &	0.0265	&	1.29	&	-2.62	&	90.52 & 1997876 \\
		Scale 1	&	0	&	0.86	&	-0.39	&	22.78	& 101881 &	0	&	0.94	&	-1.12	&	56.03 & 732366 \\
		Scale 2	&	0	&	0.60	&	-0.02	&	27.38	& 154469 &	0	&	0.68	&	-0.02	&	81.06 & 1583710 \\
		Scale 3	&	0	&	0.41	&	-0.14	&	9.49 & 10966 &	0	&	0.43	&	-0.38	&	14.05 & 31914 \\
		Scale 4	&	0	&	0.27	&	-0.27	&	9.57 & 11285 &	0	&	0.28	&	-0.36	&	14.23 & 32917 \\
		Scale 5	&	0	&	0.18	&	-0.18	&	7.13 & 4470	&	0	&	0.19	&	-0.25	&	8.09  & 6792 \\
		Scale 6	&	0	&	0.13	&	-0.28	&	5.39 & 1563 &	0	&	0.13	&	-0.29	&	5.30 & 1458 \\
		\multicolumn{11}{c}{\bf{GBP/USD}} \\
		& \multicolumn{4}{c}{Cash}  & 			\multicolumn{4}{c}{Futures}			\\			
			& Mean	& Standard	& Skewness & Kurtosis	& Jacque-Bera & Mean	 & Standard	& Skewness &	Kurtosis  & Jacque-Bera\\
			& 	& Deviation	&  &  & 	 &  & Deviation	& & &	 \\
			& \%	& \%&  & 	 &  & \%	&  \% &	 \\			\hline
		Original Data	&	0.00177	&	0.61	&	-0.16	&	6.57 & 3345	&	0.00184	&	0.64	&	-0.32	&	6.60 & 3481 \\
		Scale 1	&	0	&	0.41	&	0.09	&	5.00	& 1054 &	0	&	0.45	&	-0.04	&	4.76 & 806 \\
		Scale 2	&	0	&	0.31	&	0.00	&	6.19	& 2658 &	0	& 0.33	&	-0.02	&	5.39 & 1493 \\
		Scale 3	& 0	&	0.23	&	-0.05	&	5.93	&	2247 & 0	&	0.23	&	-0.05	& 5.32 & 1410 \\
		Scale 4	&	0	&	0.16	&	-0.17	&	6.54	& 3298 &	0	& 0.16	&	-0.14	&	5.72 & 1952 \\
		Scale 5	&	0	&	0.11	&	-0.03	&	3.84	& 185 &	0	&	0.11	&	-0.07	&	3.91 & 220 \\
		Scale 6	&	0	&	0.08	&	-0.11	&	3.56	& 92 &	0	&	0.08	&	-0.10	&	3.45 & 62 \\
			\end{tabular} \\
	\caption{\bf Descriptive statistics for log returns of Futures and Spot Series at different time-scales for Crude Oil, S\&P 500 Equity Index and GBP/USD Exchange Rate. \\ \rm Notes: The mean and standard deviation of each series is given in percentage terms, while a skewness of zero indicates no skewness and a kurtosis of 3 indicates no excess kurtosis. The Jacque-Bera statistic tests the null hypothesis that the distribution is normal and this hypothesis is rejected for all assets at all scales.}
	\label{Table1}
\end{center}
\end{sidewaystable}

%% file: Table2.tex
\begin{sidewaystable}[htbp!]
\begin{center}
	\begin{tabular}{c c || c c c c c c c c }
	Returns &   & Hedge	& Hedging	& \multicolumn{2}{c}{Standard Deviation}	& \multicolumn{2}{c}{Skewness}	& \multicolumn{2}{c}{Kurtosis} \\
Horizon	& Data Points & Ratio & Effectiveness & Unhedged & Hedged & Unhedged & Hedged & Unhedged & Hedged \\
	\hline
		Original Data & 200	& 0.91	& 0.78	& 2.43 & 1.09 &	-0.28 & -0.28	& 6.47 & 22.98 \\
		3 day & 66 &	0.95	& 0.87	& 2.98 & 1.44	& -0.32 & -0.06 & 5.53 &	13.39  \\
		6 day	& 33  & 0.99	& 0.92	& 3.23 & 1.64	& -0.32 & -0.15	& 5.29 & 14.17  \\
		12 day	& 16 &	1.00	& 0.93	& 3.32 & 1.97 &	-0.30 & 0.22 &	5.07 & 15.41 \\
	\end{tabular} \\
	\caption{\bf Hedge portfolio summary statistics, consisting of a long position in Crude Oil, hedged using futures.  \\ \rm Notes: The hedge ratio was calculated using data sub-sampled at various time-horizons. Shown are the hedge ratio and in-sample variance reduction hedging effectiveness, along with the standard deviation (in \%), skewness and kurtosis for hedge portfolio returns data, averaged over each available moving window.}
	\label{Subsampled_table}
\end{center}
\end{sidewaystable}

%% file: Table3.tex
\begin{sidewaystable}[htbp!]
\begin{center}
	\begin{tabular}{c || c | c c | c c | c c | c c}

\multicolumn{2}{c}{\bf{In-Sample}}		& \multicolumn{2}{c}{Hedging Effectiveness}	& \multicolumn{2}{c}{Standard Deviation} & \multicolumn{2}{c}{Skewness}	& \multicolumn{2}{c}{Kurtosis} \\
& Hedge Ratio & Variance & 95\% VaR HE & Unhedged & Hedged & Unhedged & Hedged & Unhedged & Hedged \\
\hline

Original Data &	0.90 &	0.75 &	0.66 &	2.40 &	1.20 &	-0.60 &	-0.31 &	12.01 &	43.05 \\
Scale 1 &	0.88 &	0.70 &	0.56 &	1.70 &	0.90 &	-0.29	 & -0.15 &	8.67 &	31.78 \\
Scale 2 &	0.89 &	0.73 &	0.62 &	1.30 &	0.60 &	-0.13 &	-0.08 &	9.30 &	29.46 \\
Scale 3 &	0.96 &	0.84 &	0.69 &	0.90 &	0.30 &	-0.15 &	-0.03 &	5.46 &	18.32 \\
Scale 4 &	0.99 &	0.94 &	0.76 &	0.60 &	0.10 &	-0.11 &	0.03 &	3.80 &	9.35 \\
Scale 5 &	1.00 &	0.98 &	0.88 &	0.30 &	0.00 &	-0.17 &	-0.11 &	3.15 &	7.65 \\
Scale 6 &	1.00 &	1.00 &	0.94 &	0.30 &	0.00 &	-0.17 &	-0.08 &	2.86 &	6.34 \\

\multicolumn{2}{c}{\bf{Out-of-Sample}}	& \multicolumn{2}{c}{Hedging Effectiveness}	& \multicolumn{2}{c}{Standard Deviation} & \multicolumn{2}{c}{Skewness}	& \multicolumn{2}{c}{Kurtosis} \\
& Hedge Ratio & Variance & 95\% VaR HE & Unhedged & Hedged & Unhedged & Hedged & Unhedged & Hedged \\
\hline
Original Data &	0.90 &	0.75 &	0.66 &	2.40 &	1.20 &	-0.60 &	-0.29 &	12.01 &	42.68 \\
Scale 1 &	0.88 &	0.69 &	0.56 &	1.70 &	0.90 &	-0.29 &	-0.13 &	8.67 &	31.52 \\
Scale 2 &	0.89 &	0.72 &	0.61 &	1.30 &	0.70 &	-0.13 &	-0.09 &	9.30 &	28.33 \\
Scale 3 &	0.96 &	0.84 &	0.68 &	0.90 &	0.30 &	-0.15 &	0.00 &	5.46 &	18.27 \\
Scale 4 &	0.99 &	0.94 &	0.76 &	0.60 &	0.10 &	-0.11 &	0.04 &	3.80 &	9.79 \\
Scale 5 &	1.00 &	0.98 &	0.88 &	0.30 &	0.00 &	-0.17 &	-0.09 &	3.15 &	7.64 \\
Scale 6 &	1.00 &	0.99 &	0.94 &	0.30 &	0.99 &	-0.17 &	-0.10 &	2.86 &	6.57 \\
	\end{tabular} \\
	\caption{\bf Statistics for Crude Oil unhedged and hedged portfolios at different scales \\ \rm Notes: Shown are the in- and out-of-sample hedge ratio, hedging effectiveness, standard deviation (in \%), skewness and kurtosis, averaged over each moving window for the unhedged and minimum-variance hedged portfolio at different scales.}
	\label{Table3}
\end{center}
\end{sidewaystable}

%% file: Table4.tex
\begin{sidewaystable}[htbp!]
\begin{center}
	\begin{tabular}{c || c | c c | c c | c c | c c}

\multicolumn{2}{c}{\bf{In-Sample}}		& \multicolumn{2}{c}{Hedging Effectiveness}	& \multicolumn{2}{c}{Standard Deviation} & \multicolumn{2}{c}{Skewness}	& \multicolumn{2}{c}{Kurtosis} \\
& Hedge Ratio & Variance & 95\% VaR HE & Unhedged & Hedged & Unhedged & Hedged & Unhedged & Hedged \\
\hline
Original Data	& 0.89	&	0.93	&	0.74	&	1.00	&	0.30	&	-0.17	&	-0.70	&	6.23	&	7.32 \\
Scale 1	&	0.86	&	0.90	&	0.68	&	0.70	&	0.20	&	0.05	&	-0.16	&	6.47	&	5.33 \\
Scale 2	&	0.90	&	0.95	&	0.78	&	0.50	&	0.10	&	0.00	&	-0.27	&	4.34	&	6.18 \\
Scale 3	&	0.94	&	0.98	&	0.85	&	0.40	&	0.10	&	-0.06	&	-0.38	&	4.54	&	6.56 \\
Scale 4	&	0.96	&	0.98	&	0.85	&	0.20	&	0.0	&	-0.03	&	-0.60	&	4.03	&	5.46 \\
Scale 5	&	0.98	&	0.99	&	0.86	&	0.20	&	0.0	&	0.17	&	-0.75	&	3.76	&	4.28 \\
Scale 6	&	0.97	&	0.98	&	0.87	&	0.10	&	0.0	&	-0.04	&	-0.03	& 3.38	&	3.17 \\

\multicolumn{2}{c}{\bf{Out-of-Sample}}	& \multicolumn{2}{c}{Hedging Effectiveness}	& \multicolumn{2}{c}{Standard Deviation} & \multicolumn{2}{c}{Skewness}	& \multicolumn{2}{c}{Kurtosis} \\
& Hedge Ratio & Variance & 95\% VaR HE & Unhedged & Hedged & Unhedged & Hedged & Unhedged & Hedged \\
\hline
Original Data	&	0.89	&	0.92	&	0.74	&	1.00	&	0.30	&	-0.17	&	-0.58	&	6.23	&	6.65 \\
Scale 1	&	0.86	&	0.89	&	0.66	&	0.70	&	0.20	&	0.05	&	-0.11	&	6.47	&	4.94 \\
Scale 2	&	0.90	&	0.94	&	0.76	&	0.50	&	0.10	&	0.00	&	-0.25	&	4.34	&	5.56 \\
Scale 3	&	0.94	&	0.97	&	0.83	&	0.40	&	0.10	&	-0.06	&	-0.29	&	4.54	&	6.31 \\
Scale 4	&	0.96	&	0.98	&	0.84	&	0.20	&	0.0	&	-0.03	&	-0.49	&	4.03	&	5.02 \\
Scale 5	&	0.98	&	0.98	&	0.86	&	0.20	&	0.0	& 0.17	&	-0.64	&	3.76	&	4.16 \\
Scale 6	&	0.97	&	0.98	&	0.87	&	0.10	&	0.0	&	-0.04	&	-0.03	&	3.38	&	2.97 \\

	\end{tabular} \\
		\caption{\bf Statistics for S\&P 500 Equity Index unhedged and hedged portfolios at different scales \\ \rm Notes: Shown are the in- and out-of-sample hedge ratio, hedging effectiveness, standard deviation (in \%), skewness and kurtosis, averaged over each moving window for the unhedged and minimum-variance hedged portfolio at different scales.}
	
	\label{Table4}
\end{center}
\end{sidewaystable}

%% file: Table5.tex
\begin{sidewaystable}[htbp!]
\begin{center}
	\begin{tabular}{c || c | c c | c c | c c | c c}

\multicolumn{2}{c}{\bf{In-Sample}}		& \multicolumn{2}{c}{Hedging Effectiveness}	& \multicolumn{2}{c}{Standard Deviation} & \multicolumn{2}{c}{Skewness}	& \multicolumn{2}{c}{Kurtosis} \\
& Hedge Ratio & Variance & 95\% VaR HE & Unhedged & Hedged & Unhedged & Hedged & Unhedged & Hedged \\
\hline
Original Data &	0.70 &	0.57 &	0.37 &	0.50 &	0.30 &	0.00 &	0.10 &	4.83 &	5.97 \\
Scale 1 &	0.55 &	0.37 &	0.24 &	0.40 &	0.30 &	0.08 &	0.05 &	3.93 &	4.69 \\
Scale 2 &	0.79 &	0.69 &	0.45 &	0.30 &	0.10 &	0.03 &	0.03 &	4.17 &	5.67 \\
Scale 3 &	0.91 &	0.88 &	0.66 &	0.20 & 0.10 &	0.01 &	0.01 &	3.72 &	5.81 \\
Scale 4 &	0.96 &	0.96 &	0.81 &	0.10 &	0.0 &	-0.10 &	0.20 &	4.55 &	5.53 \\
Scale 5 &	0.96 &	0.98 &	0.88 &	0.10 &	0.0 &	-0.12 &	0.45 &	3.48 &	4.53 \\
Scale 6	& 0.99 &	0.98 &	0.88 &	0.10 &	0.0 &	-0.07 &	0.00 &	2.89 &	4.02 \\

\multicolumn{2}{c}{\bf{Out-of-Sample}}	& \multicolumn{2}{c}{Hedging Effectiveness}	& \multicolumn{2}{c}{Standard Deviation} & \multicolumn{2}{c}{Skewness}	& \multicolumn{2}{c}{Kurtosis} \\
& Hedge Ratio & Variance & 95\% VaR HE & Unhedged & Hedged & Unhedged & Hedged & Unhedged & Hedged \\
\hline
Original Data &	0.70 &	0.56 &	0.37 &	0.50 &	0.30 &	0.00 &	0.09 &	4.83 &	5.98 \\
Scale 1 &	0.55 &	0.35 &	0.23 &	0.40 &	0.30 &	0.08 &	0.05 &	3.93 &	4.85 \\
Scale 2 &	0.79 &	0.69 &	0.45 &	0.30 &	0.10 &	0.03 &	0.02 &	4.17 &	5.65 \\
Scale 3 &	0.91 &	0.88 &	0.66 &	0.20 &	0.10 &	0.01 &	0.00 &	3.72 &	5.73 \\
Scale 4 &	0.96 &	0.96 &	0.80 &	0.10 &	0.0 &	-0.10 &	0.17 &	4.55 &	5.51 \\
Scale 5 &	0.96 &	0.98 &	0.87 &	0.10 &	0.0	& -0.12 &	0.43 &	3.48 &	4.48 \\
Scale 6 &	0.99 &	0.98 &	0.88 &	0.10 &	0.0 &	-0.07 &	-0.03 &	2.89 &	3.97 \\

	\end{tabular} \\
			\caption{\bf Statistics for GBP/EUR Exchange Rate unhedged and hedged portfolios at different scales \\ \rm Notes: Shown are the in- and out-of-sample hedge ratio, hedging effectiveness, standard deviation (in \%), skewness and kurtosis, averaged over each moving window for the unhedged and minimum-variance hedged portfolio at different scales.}
	\label{Table5}
\end{center}
\end{sidewaystable}